\documentstyle[12pt]{article}
\begin{document}
\begin{center}
~~
~~

~

\noindent
{\Large \bf On the interaction of extreme-low-frequency (ELF) radiation 
with living matter's coherent spiral states}             
\vskip1.0cm

{\it  K. ZIOUTAS}
\vskip1cm
{\it Physics Department, University of Thessaloniki,    \\
  GR-54006  Thessaloniki,  Greece.         }
\vskip1.0cm
Email :~~ZIOUTAS@CERNVM.CERN.CH

\vskip1.0cm
{\it 5. 1. 1996}
\vskip1.5cm
{\bf Abstract.}
\end{center}  

\noindent
{\bf  Conventional physics cannot explain bioeffects caused by  
ELF fields \cite{adair}. A (double) resonance interaction 
of ELF radiation with the  highly sub-thermal coherent motion of spiral 
waves  in living matter (LM) is suggested. The (geo)magnetic 
cyclotron resonance (CR) absorption  by the constituent 
ions can drastically change their non-thermal degrees of freedom 
in the spiral state, destroying their pattern, which is heuristically 
assumed to be part of the cell signaling. Combining ELF irradiation with 
calcium imaging to unravel the suggested mechanism, a new instrument 
in biomedicine might emerge.  These fascinating waves, being part of the 
bio-machinery, might provide insight in cancer, embryogenesis, etc., 
as they are affected by ELF radiation in an as yet unknown manner [2-6]. 
}

\newpage

\section*{1. Introduction}

There is a general agreement that the ubiquitous in modern civilization
nonionizing ELF electromagnetic fields induce a variety of
physiological and cellular responses 
(e.g. [4$-$10]).
Concerns have been raised that they are carcinogenic
and leukegenic \cite{adair,kwee}. The underlying interaction with 
LM is considered to be too weak for us to expect a significant effect
at the cellular level, since such effects are usually associated with
the breaking of molecular bonds of macromolecules like DNA \cite{adair}.
The characteristic quantum energy $h\nu_{ELF} \approx 10^{-14}$ eV
of the ELF-radiation (e.g. 50 or 60 Hz) is far too below the thermal energy
($kT \approx 2.4\cdot 10^{-2}$ eV), or, much worse, the chemical binding 
energy (a few eV), to disrupt the genetic code. 
The thermal noise fields, in the vicinity of cells, are~ $\sim 100-1000$ times
stronger than that associated with the  external ELF field \cite{lito},
which is composed, in the classical picture, of quasi-independent electric 
and magnetic field components. Therefore, the question remains, how do cells 
recognize the existence of the primary weak signal ? \cite{krug}.
One arrived to the conclusion that a fundamental understanding of the 
(as yet unknown [2$-$6])
interaction mechanism of ELF fields inducing biological effects must be
found {\sf outside conventional physics} \cite{adair}. That is to say, the most natural 
conclusion is that there cannot be biological actions by weak ELF fields
\cite{krug,bennet},  contradicting experimental results showing their impact
on DNA, RNA and protein synthesis, $Ca^{++}$ regulation, 
dynamics of cell division, embryogenesis, etc. 
[13$-$17],
including biomineralization in vitro \cite{dalas}.

Lowest external perturbations may trigger bioeffects, as
LM functions under conditions far from equilibrium 
\cite{grundler,nicolis}. 
Some proposals to solve this signal-to-noise problem are mentioned :
~{\bf a)}~ stochastic resonance at the driving frequency and
its harmonics, i.e. weak signal amplification by the system noise itself
\cite{krug}, ~{\bf b)}~ cells average out the thermal noise by integrating
the  signals, with the estimated integration time being 
$\sim 10$ times longer than exposure intervals observed to produce
bioeffects \cite{weaver}, ~{\bf c)}~ cells discriminate against the spatially
random thermal noise fields by recognizing them (somehow) as spatially
incoherent \cite{lito} : the data show that a (temporally coherent) 100 Hz
field causes abnormal embryogenesis, being blocked by a
superimposed noise field ($\sim 10-400$ Hz), and ~{\bf d)}~ the weak fields 
couple to receptor-controlled cytosolic calcium oscillator, being stabilized
far away from thermal equilibrium, predicting specific intracellular calcium 
oscillations \cite{grundler}.~~~~This list of ideas is by no means 
exhaustive (e.g. \cite{frey}).

In this work an interaction mechanism between ELF radiation and LM is
described, which can be directly established combining known
experimental methods, introducing possibly a new technique in biomedicine.

\section*{2. Spiral waves}  

Excitable media in (bio)chemistry being initially in a spatially uniform 
steady state can reach and maintain spontaneously a far-from-equilibrium 
non-uniform 2D or 3D spatiotemporal order \cite{winfree}; 
they are characterized by their
ability to propagate signals undamped over long distances at the expense
of (chemical) energy stored in the medium. 
A rotating spiral wave \cite{winfree}, with inward
turning tip (inner endpoint) and outward moving fronts, is the 2D 
cross section through the real 3D scroll wave, which emanates from an 
organizing closed axis. 

In biology, rotating spiral waves have been observed in numerous cases 
[22$-$25] :
in retinal and cortical nerve nets, in heart and muscular tissue, in the 
lenses of the eye of a firefly, but also in the aggregation of cells
\cite{nicolis,gerisch}, 
as well as, in fungi culture growth and sporulation (with circadian or 
sub-circadian biorhythms) \cite{bouret}.
$Ca^{++}$ wave development has been observed even at the 
level of a single cell \cite{leich}. Calcium  couples
extracellular stimuli to cellular responses in virtually all cell types
\cite{snyder}; the most obvious function of calcium waves is to carry 
calcium signals deep into the cells \cite{jaffe}.
However, the exact mechanism of $Ca^{++}$ signaling that
mediates cell communication remains one of the most intriguing mysteries
in biology \cite{clapham}. 
The frequency and amplitude of the propagating $Ca^{++}$ waves (remarkably
constant for individual cells) potentially contains additional encoded
information \cite{clapham,sneyd,berridge}. Furthermore, 
cell-specific unique patterns of transient 
$Ca^{++}$ signals have been termed 
`$Ca^{++}$ {\it fingerprint'}, 
being possibly implicated in complex processes such as spatial cellular
specialization, cellular differentiation, and information storage
\cite{berridge,prent}.
A $Ca^{++}$ transient, which initiates development during fertilization,
takes also the form of wave or oscillation \cite{galione}.
                           
Typical values for the velocity of propagation, wavelength and 
frequency  of spiral waves in LM,
are $v \approx 10-100~\mu$m/s, $\lambda \approx 10-1000~\mu$m and
$\nu \approx 0.1-10$ Hz, respectively (e.g. \cite{jaffe,clapham,lipp,david}). 
The resonance response of spiral waves 
to the {\it modulation frequency} of an external stimulus
and its harmonics \cite{res} can in addition enhance the interaction
with varying weakest irradiation.

\section*{3. The suggested interaction mechanism}

Inorganic ions like $Ca^{++},~K^+,~Na^+,~Mg^{++},~Cl^-$, etc. make $\sim 1$
$\%$ of the total cell weight.
Ion transport is related to cell signaling in many ways \cite{garcia}.
Throughout this work for reasons of simplicity, numerical 
examples refer to ionized calcium, which is the most common signal
transduction element in cells ranging from bacteria to specialized
neurons \cite{clapham}.
Spiral waves of this ubiquitous second messenger ($Ca^{++}$) in LM
are of particular interest, since their collective ($\approx$ coherent)
rotation frequency is within the range of the ion's CR frequency 
\cite{zi} :
\begin{equation}
\omega_{cyclotron} = 2\pi \nu_{cyclotron} = 
\frac{qB}{M} \approx 40~Hz 
\end{equation}
for $Ca^{++}$ being inside the static (geo)magnetic field 
$B \approx 0.5$ gauss (M = ion mass).
The excessive large orbit of the gyrating ions and the 
collisions with the surrounding medium \cite{adair} make
a classical discription of CR meaningless. In the quantum picture 
\cite{zi}, considering the excitation of quasi-quantum states 
(= Landau Levels), the photon energy is transformed first into kinetic 
energy of the gyrating ion; those long-lived states \cite{zi} are actually 
not deexcited radiatively, as they give their energy to the environment
during collisions.

\noindent
{\sf The gyroresonance} frequency (relation (1)) 
refers to an ideal case, with the incident photons moving parallel 
to the guide magnetic field ($B$), otherwise
only the parallel field component contributes. Inside the ubiquitous 
static geomagnetic field, LM  is exposed to a rather isotropic 
ELF irradiation coming from the  power lines, but also from geomagnetic 
activity \cite{hahn}.
Relation (1) provides the upper limit of the fundamental frequency. 
For example : {\bf a)} the measured width of the gyroresonance signal,
e.g. the variation of the cell-motility as a function of the frequency of
the incident electromagnetic radiation \cite{liboff}, is a wide Lorentzian 
shaped distribution at the expected CR frequency of  $Ca^{++}$ at
$\omega_{cyclotron}$ = 16~Hz
for $B=0.209$ gauss, with FWHM $\approx 6$ Hz, 
{\bf b)} a weak 16 Hz irradiation in combination with a 0.234
gauss parallel static field, i.e. also 
at the $Ca^{++}$ gyroresonance frequency,
provided an inhibition of calcium influx in thymocytes \cite{liburdy}, and
{\bf c)} an enhancement of $Ca^{++}$ uptake by normal and malignant 
lymphocytes was observed at 13.6 Hz (near the CR for $Ca^{++}$) 
\cite{lyle}.
Thus, the gyroresonance interaction and its higher harmonics with 
the $Ca^{++}$ ions alone inside the geomagnetic field ($\approx$ 0.5 gauss)
can occur actually at any frequency below $\sim 10-100$ Hz.
However, energy transfer from the weak ELF radiation to LM via the
CR absorption was also considered to be negligibly small, as the thermal
Brownian-like motion overwhelms by factor $\sim 10^{12}$ any (orbital)
gyromotion \cite{adair}.
Afterall, the CR interaction probability reaches unitarity, as the 
resonance cross section in the classical or quantum picture is 
enormously high :
~$\sigma^{CR} \approx \pi \lambda^2_{ELF}$~ \cite{zi}.
Thus, the CR interaction alone provides primarily the necessary  mechanism 
to efficiently couple the ELF radiation to LM's ions.

\noindent
{\sf The spiral wave} spinning frequency in LM  is,
surprisingly, within the geomagnetic CR frequency range of the constituent  
ions. Moreover, it occasionally occurs that the ELF field is simultaneously 
also at resonance with the spiral coherent macroscopic rotation.
Although such a potential double resonance can further enhance the
interaction \cite{res}, 
its occurence is not absolutely necessary (s. below). 
The fundamental question to be addressed is how this coupling overcomes the
very unfavourable thermal energy.
Note that the propagation of $Ca^{++}$ waves is an active process, not solely the result 
of passive calcium diffusion \cite{sylvain}.
Along with ref. \cite{grundler}, those degrees of freedom which govern
these spiral states must be at least partially decoupled from the rest 
of the system, which might well be in thermodynamic equilibrium. 
Only then, the following amazing numerical estimations, or conclusions, 
make sense (relations (2)$-$(4)).
Remarkably, inside an environment with 
a mean thermal energy~ kT $\approx 10^{-2}$ eV~  per atom or molecule or ion, 
the constituent ions of the coherent spiral motion appear to 
propagate at an almost constant velocity ($v$) taken to be typically 
\cite{jaffe}
\begin{equation}
v = v_{spiral} \approx 50~\mu m/s,
\end{equation}
which should correspond to a kinetic energy of a $Ca^{++}$ ion following the  
collective spiral motion of
\begin{equation}
T^{Ca}_{kin} = \frac{1}{2} M v^2 \approx 0.5\cdot 10^{-15}~eV .
\end{equation}
This means that the spiral-ions keep a coherent velocity 
component ($v=v_{spiral}$), which is slower 
than their thermal velocity in the surrounding medium
($v_{th} \approx 3.5\cdot 10^4$ cm/s), by a factor $\sim 10^7$.
To put it differently, it is as if one could assign to the spiral's 
constituents an {\it effective temperature} 
($kT^{effective} \approx T^{Ca}_{kin}$),
being {\it far below} the thermal equilibrium:
\begin{equation}
T^{effective} \approx 10^{-10}-10^{-13}~K
\end{equation}
for ~$v \approx (5-200)~\mu$m/s \cite{jaffe}, while being inside 
an environment approximately at room temperature ($!$) .
Moreover, due to the peculiar spiral motion, the participating ions 
appear finally not to move randomly
relative to each other, since their velocity is 
highly correlated, as they are driven by the common spiral wave.
Bearing in mind that cryogenic thermal detectors in dark matter search,
work in the $\sim mK$ range \cite{minowa}, one should expect a very high 
sensitivity of this form of macroscopic `frozen' states inside LM 
to external signals.

Let us consider the CR absorption of one single ELF radiation quantum by 
a $Ca^{++}$ ion belonging  to a spiral wave,
taken the earth's magnetic field as the guide field for the
CR interaction. 
One single photon energy ($\approx 10^{-14}$ eV), being transformed 
to gyromotion (= kinetic energy) of an ion, changes obviously 
its coherent motion and the associated
kinetic energy ($\approx 0.5\cdot 10^{-15}$ eV) completely.
This is still true even for initially quite higher $Ca^{++}$ velocities,
e.g. $v \approx 200~\mu$m/s. 
Furthermore, assuming a much higher spiral velocity, 
or, a much weaker guide magnetic field for the CR to occur,
the resonance absorption of several ELF field 
quanta per ion can take place because of the enormous high CR cross section
involved; therefore, they also can change finally the ion's collective
velocity, disturbing or even destroying the fascinating biochemical 
spatiotemporal order.
This hypothetical case shows that the spiral wave can be 
disturbed, even if its intrinsic frequency
is not in resonance  with the ELF field; 
of course, if it happens to be the case probably 
the impact due to the double resonance will be even stronger.

\underline{\it In short} : only spiral states, stabilized (= `frozen') 
far below the thermal equilibrium,
can drastically be changed by the weakest ELF~ CR photon absorption,
disturbing thus this form of signal transmission,
provided the wave's coherent degrees of freedom are decoupled
from the thermal ones. With spirals being involved in LM's fine tuned 
machinery, and, having in mind the
plethora of cellular events $Ca^{++}$ controls (e.g. proliferation of many 
cell types \cite{lyle,means}), it is not
unreasonable to expect all kinds of biological malfunction,
once its pattern has been externally modified.

\section*{4. Discussion - Suggestion}

The spontaneous appearance in LM of spatiotemporally ordered states, 
which move collectively as spinning and occasionally as drifting 
spirals with  characteristic rotation frequency and at a very slow 
(not random) velocity of propagation, distinguishes all excitable 
media in the living and the non-living world.
This work combines the known properties of the fascinating spiral waves,
which dominate biology, with the (geomagnetic) 
cyclotron resonance interaction
of the constituent ions without inventing necessarily new physics.
The observed impact of ELF radiation on the calcium metabolism of the
cell may be due to changes caused in the calcium waves involved.
It has been recognized already how important unraveling the multiple 
roles played by $Ca^{++}$ will be, e.g. in regulation of cell proliferation 
\cite{means}.

In conclusion the following suggestions are underlined: ~ {\bf a)} 
~there are (highly developed) experimental techniques to make  
the heuristically assumed response of the $Ca^{++}$ waves to ELF radiation, 
or to modulated electromagnetic irradiation at the spiral's own 
frequency, `visible' ;  this can be achieved, for example, by
the use of fluorescence imaging and confocal microscopy 
with $\sim \mu$m and $\sim$ ms, space and time resolution, respectively 
\cite{jaffe,lipp,gillot,vmf,lippi}.
Such a direct observation, being the paramount component of this work, 
could reconcile the mystery surrounding the 
connection between ELF fields and the biological observations,
establishing thus the role of spiral waves in biocommunication, ~{\bf b)}~
via the  ELF irradiation of LM, one should also search for
any difference between normal and malignant cells, which could be utilized
to selectively interact externally with the cancer cells only, and {\bf c)}~
the spiral multicellular morphogenesis may be considered a very primitive
form of embryogenesis, where cell differentiation might be associated
with enhanced signaling. Bearing in mind the observed abnormalities due 
to ELF irradiation during embryogenesis \cite{ubeda}, another direct 
piece of evidence of the impact of ELF radiation on spiral waves could
be derived from the study of the multicellular aggregation 
\cite{gerisch}, or fungi culture growth \cite{bouret}, under ELF irradiation.
The striking macroscopic spirals appearing in those investigations simplify
the experiment enormously.

Thus, the ELF irradiation in combination with existing biological methods 
may become a new probe, providing possibly  access to 
microscopic biological phenomena. One may finally gain insight into open
issues like embryogenesis, cell signaling and differentiation, 
biorhythmicity, etc. . Similar experiments performed with excitable media
from the non-living world, can be helpful, as spiral waves are strongly 
affected by just a few V/cm static electric fields \cite{muller}.


\newpage



\begin{thebibliography}{99}


\bibitem{adair} R. K. Adair,   Phys. Rev. {\bf A43} (1991) 1039. 
 s. also D. P. Hamilton,  Science {\bf 251} (1991) 863.
\bibitem{krug}  I. L. Kruglikov, H. Dertinger,  Bioelectromagn.
 {\bf 15} (1994) 539. 
\bibitem{garcia} J. Garcia-Sancho, et al., Bioelectromagn. 
 {\bf 15} (1994) 579.
\bibitem{martin} A. H. Martin,  Bioelectromagn. {\bf 13} (1992) 223.
\bibitem{good}  E. M. Goodman, B. Greenebaum, M. T. Marron, Bioelectromagn. 
 {\bf 15} (1994) 77.
\bibitem{lito} T. A. Litowitz, et al.,  Bioelectromagn. 
 {\bf 15} (1994) 105, and references therein.
\bibitem{weis} D. Weisbrot, et al.,  Bioelectrochem. Bioenerg. {\bf 31}
 (1993) 39.
\bibitem{mich} S. M. Michaelson,  Health Phys. {\bf 61} (1991) 3. 
\bibitem{anders} L. E. Anderson,  Health Phys. {\bf 61} (1991) 41.
\bibitem{berg} H. Berg,  Bioelectrochem. Bioenerg. {\bf 31} (1993) 1.
\bibitem{kwee} S. Kwee, P. Raskmark,  Bioelectrochem. Bioenerg. 
 {\bf 36} (1995) 109.
\bibitem{bennet} W. R. Bennet, Jr.,  Phys. Today {\bf 47}(4) (1994) 23.
\bibitem{alipov} Ye. D. Alipov, I. Ya. Belyaev, O. A. Aizenberg, 
 Bioelectrochem. Bioenerg. {\bf 34} (1994) 5.
\bibitem{hendee} W. R. Hendee, J. C. Boteler,  Health Phys. 
 {\bf 66} (1994) 127, and references therein.
\bibitem{grundler} W. Grundler, et al.,  Naturwissenschaften {\bf 79}
 (1992) 551; C. Eichwald, F. Kaiser,  Biophys. J. {\bf 65} (1993) 2047, 
 Bioelectromagn. {\bf 16} (1995) 75.
\bibitem{ubeda} A. Ubeda, et al., Bioelectromagn. {\bf 15} (1994) 385.
\bibitem{ernst} E. Niggli, P. Lipp,  Cardiovasc. Res. {\bf 29} (1995) 441.
\bibitem{dalas} E. Dalas, D. Fatouros,  J. Cryst. Growth {\bf 125} (1992) 27.
\bibitem{nicolis}  G. Nicolis, I. Prigogine,  Exploring Complexity,
 W. H. Freeman $\&$ Co, New York (1989) 31-36.
\bibitem{weaver} J. C. Weaver, R. D. Astumian, Science {\bf 247} (1990) 459.
\bibitem{frey} A. H. Frey,  FASEB J. {\bf 7} (1993) 272.
\bibitem{winfree} For example : 
 A. T. Winfree, S. H. Strogatz,  Physica {\bf D8} (1983) 35, 
 Nature {\bf 311} (1984) 611;
 A. T. Winfree,  Physica {\bf D12} (1984) 321,
 SIAM Rev. {\bf 32} (1990) 1, Nature {\bf 371} (1994) 233.
\bibitem{aranson} I. Aranson, D. Kessler, I. Mitkov,  Physica {\bf D85}
 (1995) 142, and references therein.
\bibitem{pertsov} A. Pertsov, M. Vinson,  Phil. Trans. R. Soc. London 
 {\bf A347} (1994) 687.
\bibitem{madone} B. F. Madone, W. L. Freedman,  Am. Sci. {\bf 75} (1987) 252.
\bibitem{gerisch} G. Gerisch, Naturwissenschaften {\bf 58} (1971) 430.
\bibitem{bouret} J. A. Bouret, R. G. Lincoln, B. H. Carpenter, Science 
 {\bf 166} (1969) 763.
\bibitem{leich} J. D. Leichleiter, D. E. Clapham,  Cell {\bf 69} (1992) 283.
\bibitem{snyder} P. M. Snyder, K.-H. Krause, M. J. Welsh, J. Biol. Chem. 
{\bf 263} (1988) 11048.
\bibitem{jaffe} L. F. Jaffe,  Proc. Natl. Acad. Sci. USA {\bf 88} (1991) 9883.
\bibitem{clapham} D. E. Clapham,  Cell {\bf 80} (1995) 259, 
 Nature {\bf 375} (1995) 634, {\it ibid.} {\bf 364} (1993) 763.
\bibitem{sneyd} J. Sneyd, A. C. Charles, M. J. Sanderson, Am. J. Physiol.
 {\bf 266} (1994) C293.
\bibitem{berridge} M. J. Berridge, R. F. Irvine, Nature {\bf 341} (1989) 197.
\bibitem{prent} M. Prentki, et al.,  J. Biol. Chem. {\bf 263} (1988) 11044. 
\bibitem{galione} A. Galione, et al.,  Science {\bf 261} (1993) 348,
 H. C. Lee, R. Aarhus, T. F. Walseth, {\it ibid.} (1993) 352.
\bibitem{lipp} P. Lipp, E. Niggli,  Circ. Res. {\bf 74} (1994) 979;
 J. M. Davidenko, P. Kent, J. Jalife,  Physica {\bf D49} (1991) 182.
\bibitem{david} J. M. Davidenko, et al.,  Nature {\bf 355} (1992) 349, 
 P. Camacho, J. D. Leichleiter, Science {\bf 260} (1993) 226.
\bibitem{res} M. B\"ar, et al.,  Phys. Rev. Lett. {\bf 74} (1995) 1246,
 M. Braune, A. Schrader, H. Engel,  Chem. Phys. Lett. {\bf 222} (1994) 358 
 and Phys. Rev. {\bf E52} (1995) 98, 
 O. Steinbock, V. Zykov, S. C. M\"uller, Nature {\bf 366} (1992) 322.
\bibitem{zi} K. Zioutas,  Phys. Lett. {\bf A189} (1994) 460. 
\bibitem{hahn} O. Hahneiser, et al.,  Bioelectrochem. Bioenerg. {\bf 37}
 (1995) 51.
\bibitem{liboff} A. R. Liboff, et al.,  {\it in} ELF electromagnetic fields:
 The question of cancer, Edits: B. W. Wilson et al., Battelle Press, 
 (1990) 251; B. R. McLeod, et al.,  J. Biolectricity {\bf 6} (1987) 1; 
 s. also discussions in : A. V. Prasad, et al.,  Health Phys. {\bf 66} 
 (1994) 305, L. A. Coulton, A. T. Barker,  Phys. Med. Biol. 
 {\bf 38} (1993) 347.
\bibitem{liburdy} R. P. Liburdy,  FEBS Lett. {\bf 301} (1992) 53; 
 M. G. Yost, R. P. Liburdy,  {\it ibid.} {\bf 296} (1992) 117.
\bibitem{lyle} D. B. Lyle, et al.,  Bioelectromagn. {\bf 12} (1991) 145;
 S. M. Ross, {\it ibid.} {\bf 11} (1990) 27.
\bibitem{sylvain} S. DeLisle, M. J. Welsh,  J. Biol. Chem. {\bf 267}
 (1992) 7963.
\bibitem{minowa} E.g. :  M. Minowa, et al.,  
 Nucl. Instr. Meth. in Phys. Res. {\bf A327} (1993) 612.
\bibitem{means} A. R. Means,  FEBS Lett. {\bf 347} (1994) 1.
\bibitem{gillot} I. Gillot, M. Whitaker,   J. exp. Biol. {\bf 184} (1993) 213.
\bibitem{vmf} V. M. Fernades de Lima, M. Goldermann, W. R. L. Hanke,  
 Brain Res. {\bf 663} (1994) 77.
\bibitem{lippi} P. Lipp, E. Niggli,  Biophys. J. {\bf 65} (1993) 2272.
\bibitem{muller} J. Sch\"utze, O. Steinbock, S. C. M\"uller,  
 Nature {\bf 356} (1992) 45;  O. Steinbock, J. Sch\"utze, S. C. M\"uller,   
 Phys. Rev. Lett. {\bf 68} (1992) 248; 
 A. P. Manuzuri, et al.,  Phys. Rev. {\bf E48} (1993) 3232.
                                     

\end{thebibliography}
\end{document}